\documentclass{sig-alternate-10pt}

\usepackage{setspace}
\usepackage[T1]{fontenc}
\usepackage{float}
\usepackage{amssymb}
\usepackage{amsmath}
\usepackage{amsfonts}
\usepackage{amssymb}
\usepackage{algorithmic}
\usepackage{algorithm}

\usepackage{color}
\usepackage{soul}
\usepackage[hyphens]{url}

\graphicspath{{.}{./figs/}}

\begin{document}

\title{Nash-Peering: A New Techno-Economic Framework for Internet Interconnections}

\numberofauthors{4}
\author{
\alignauthor Doron Zarchy\\
\affaddr{Hebrew University of Jerusalem}\\
\email {\tt doronz@cs.huji.ac.il}
\alignauthor Amogh Dhamdhere\\
\affaddr{CAIDA}\\
\email {\tt amogh@caida.org}
\and \alignauthor Constantine Dovrolis\\
\affaddr{Georgia Tech}\\
\email {\tt constantine@gatech.edu}
\alignauthor Michael Schapira\\
\affaddr{Hebrew University of Jerusalem}\\
\email{\tt schapiram@huji.ac.il}
}

\maketitle

\begin{abstract}
The current framework of Internet interconnections, based on transit and settlement-free peering relations, has systemic problems that often cause peering disputes. We propose a new techno-economic interconnection framework called Nash-Peering, which is based on the principles of Nash Bargaining in game theory and economics. Nash-Peering constitutes a radical departure from current interconnection practices, providing a broader and more economically efficient set of interdomain relations. In particular, the direction of payment is not determined by the direction of traffic or by rigid customer-provider relationships but based on which AS benefits more from the interconnection. We argue that Nash-Peering can address the root cause of various types of peering disputes.
\end{abstract}

\section{Introduction} \label{sec:introduction}
The tussle between content and access providers has led to interest from techologists 
and policy makers in the bargaining,
money flow, and market power issues behind Internet interconnections \cite{fcc15}. 
Peering disputes over traffic
imbalances are not new -- several such incidents between large
ISPs and content providers have happened over the years \cite{bafna:2014}.
More recently however, such disputes have been fueled by exploding demand for streaming video, and
growing concentration of content among a few providers and CDNs,
raising questions about appropriate network management, 
interconnection business strategies, 
and the impact of these peering disputes on end-user performance \cite{clark2014measurement}.  

An important point about the aforementioned peering disputes
is that, while they result in performance degradation, they are mostly
{\em economic issues}. The technical solutions are straightforward --
increase the capacity of the network (or shift traffic to other
routes) as demand increases. The key questions,
however, relate to which party should pay for infrastructure upgrades
and how the costs and benefits of interdomain relationships should be
split among the different parties.

The current interconnection framework is predominantly based
on two types of bilateral relations between Autonomous Systems (ASes): 
{\em transit} and settlement-free Peering ({\em sf-Peering}) \cite{huston:1998}.
In transit relations, the provider advertises global Internet routes to the customer, while
the latter pays based on the traffic that it sends/receives.
In sf-Peering, there is no exchange of money and each AS only exports
routes that originate from itself and from its customer-cone.
The conditions for establishing an sf-Peering relation depend on the business
profile of the corresponding ASes. For instance, content providers tend to peer
openly (to minimize their transit fees), while transit providers are
more selective, aiming to not peer with ASes that may become their customers in 
the future \cite{lodhi:2014}. 
Other types of relations include {\em paid-peering} (similar to sf-Peering but one
of the two peers gets a traffic-dependent payment from the other)
and {\em partial transit} (where only some routes are advertised from the provider
to the customer) \cite{faratin:08}. 

Our main premise is that the currently deployed interconnection framework has fundamental weaknesses and
systemic problems that will continue causing peering disputes.
One such weakness is that sf-Peering relationships are formed for reasons 
(e.g., a balanced traffic ratio) that are often only indirectly related to 
the costs and benefits of the interconnection \cite{norton-peering-playbook}. 
Consequently, some interconnections may not be formed even when they 
can result in both bilateral and Internet-wide benefits. 
Instead of looking at each peering conflict as an isolated incident,
we need to focus on the limitations of the interconnection
framework that is generating these disputes.

In this paper, we propose a new interconnection paradigm that we refer to as 
\emph{Nash-Peering} because it is based on the principles of Nash
Bargaining in game theory and economics. Nash-Peering constitutes a
radical departure from today's interconnection practices, providing a
broader and more economically efficient set of interdomain relations.
In particular, a Nash-Peering interconnection is established if and only if it is beneficial for
both parties.
Additionally, the direction of payment is not determined by the direction of traffic
or by rigid customer-provider relationships 
but based on which AS benefits more from the interconnection. 
We argue that Nash-Peering can address the root cause of various types of peering disputes. 
We finally discuss how the Internet can transition to Nash-Peering.

\section{Nash-Peering} \label{sec:proposed}

\subsection{Overview of Nash bargaining}
Alice and Bob negotiate how to split a hidden treasure of value $V$. 
Each of them has some information about the treasure's location and they can only find it if they 
agree on how to split $V$.  
To recover the treasure each of them will have to spend all their savings: 
Alice has $s_A$ and Bob has $s_B$.
Obviously, they will be interested in the treasure only if $s_A+s_B<V$.
If they agree, each of them will get a share, say $v_A$ and $v_B$ respectively, with $v_A+v_B=V$.
Otherwise, if they do not reach agreement, they will stay with their savings $s_A$ and $s_B$;
this is their ``outside option'' (also referred to as disagreement or threat point).  

If they reach an agreement, their cumulative benefit or {\em surplus} is $\Delta=V-(s_A+s_B)$.
The Nash Bargaining solution postulates that Alice should get her outside option $s_A$ plus half the surplus;
the other half should go to Bob.  In other words, 
Alice should get $v_A=s_A+\Delta/2=(V+s_A-s_B)/2$, while Bob should get $v_B=s_B+\Delta/2=(V+s_B-s_A)/2$.
Note that if Alice, for instance, invests more than Bob ($s_A > s_B$), she should get a larger portion
of the treasure, which is arguably a more fair allocation than just splitting $V$ in two equal shares.

Nash showed in 1950 that the previous allocation follows mathematically from a set of reasonable 
assumptions about the rationality of the two involved individuals \cite{nash1950bargaining,nash1953two}; this is 
an axiomatic theory because it does not derive the Nash solution based on a model of the actual 
bargaining process.
The Nash solution is Pareto-optimal and fair in the sense that it splits the surplus equally between
the two players.\footnote{There are also other notions of fair bargaining, such as Kalai's solution
that maximizes the minimum payoff among the two players.} 
In 1986, Binmore, Rubinstein and Wolinsky modeled the bargaining process between two players as a
sequential non-cooperative game, and they proved, under certain assumptions, that the unique 
equilibrium is actually the previous Nash solution $(v_A,v_B)$
(namely, each player gets her outside option plus half surplus) \cite{binmore1986nash}.
An important condition behind this result is that the risk of breakdown 
(i.e., the probability that Alice or Bob will walk away from the negotiations even though $s_A+s_B<V$)
is the same for both players and negligible. 

\subsection{Formulation}
Suppose that two ASes, A and B, consider having a direct interconnection between them. 
To argue quantitatively about the costs and benefits of this interconnection, A and B need to be
very specific about the routes and traffic volume that will be 
going through this interconnection as well as about its location and capacity.  
Suppose for now that A and B are only negotiating for an interconnection at a specific exchange point 
at which they are both present, and for a single BGP route that A will export to B
advertising reachability to a destination prefix D.  
Suppose that this traffic volume is denoted by $T$. 
In the rest of this section, all costs and payments refer only to $T$.
We will consider the more general case in Section~\ref{sec:practice}. 

In the absence of a direct interconnection, flow $T$ is routed through a different network path. 
For instance, $T$ could flow from B to a transit provider C and then to A. 
This is the ``outside option'' of A and B.
Let $v'_A$ and $v'_B$ be the payoffs of A and B, respectively, when they are not directly
interconnected. 
If $c'_A$ is the cost incurred by A for flow $T$ in the outside option, we have that $v'_A = -c'_A$;
similarly for B, $v'_B = -c'_B$.
The costs $c'_A$ and $c'_B$ may represent, for instance, transit fees that A and B need to pay to 
their common transit provider C. 
In a more general case, in which the outside option is associated with both costs and revenues, 
$c'_A$ or $c'_B$ represent the net ``costs minus revenues'' amount and they may even be negative.

If A and B reach an agreement, $T$ will flow directly from B to A. In general, 
there will be some costs $c_A$ and $c_B$, for A and B respectively, associated with this flow. 
These terms can represent the internal costs of A and B for carrying 
this traffic through their own infrastructure. 

The surplus of the direct interconnection is 
\begin{equation}
\Delta = (c_A' + c'_B) - (c_A+c_B)	\label{eq:surplus}
\end{equation}
Obviously the interconnection makes sense only if the surplus is positive, $\Delta > 0$.
According to the Nash solution, $\Delta$ should be equally split between A and B.
In the general case, this can only happen if there is a payment between the two players. 
To see that, consider the case $c'_B - c_B > c'_A - c_A$, i.e., B benefits from 
the interconnection more than A. If so, the surplus can be split equally between A and B if B pays 
a fee $r>0$ to A, such that 
\begin{equation}
r = \frac{(c'_B - c_B) - (c'_A - c_A)}{2} \label{eq:nash-price}
\end{equation}
In that case, the benefit that B gets from the interconnection is  equal to the benefit that A
gets from the interconnection because $c'_B - c_B - r = c'_A - c_A + r$.
Then, the payoff of A will be 
\begin{equation}
v_A = r - c_A = \frac{(c'_B - c'_A) - (c_A+c_B)}{2} = v'_A + \frac{\Delta}{2} \label{eq:utilityA} 
\end{equation} 
and the payoff of B will be  
\begin{equation}
v_B = - r - c_B = \frac{(c'_A - c'_B) - (c_A+c_B)}{2} = v'_B + \frac{\Delta}{2} \label{eq:utilityB} 
\end{equation} 
Note that $(v_A,v_B)$ is the Nash solution. 

It is easy to see that {\em the payoff of each AS at the Nash solution ($v_A$ and $v_B$) is higher than its payoff 
at the outside option ($v'_A$ and $v'_B$, respectively) if and only if $\Delta>0$.}  
So, with Nash-Peering both ASes are better off compared to the outside option, even if one of
them has to pay the other.  

If A benefits more from the interconnection than B ($c'_B - c_B < c'_A - c_A$), the 
previous equations still hold but for a negative payment $r$, i.e., A would need to pay $-r$ to B 
to obtain the Nash solution.  

The previous simple equations describe the proposed {\em Nash-Peering} interconnection framework. 
To summarize, A and B should agree to directly interconnect if and only if the surplus 
$\Delta$ is positive because in that case both of them will benefit. 
If B benefits more from the interconnection than A then B should pay $r$ to A, 
where $r$ is given by (\ref{eq:nash-price});
otherwise A should pay $-r$ to B. 
It is important to note that the Nash-Peering framework does not rely on any arbitrary conventions
about the direction of payment; it does not assume that the sender of the traffic should always pay, for instance.

\subsection{Remarks}
The negotiation terms involve only the costs incurred by A and B when the 
interconnection is not in place ($c'_A$ and $c'_B$) and when it is ($c_A$ and $c_B$).
These costs depend, in general, on the location of the interconnection. 
For example, if A is a US-based ISP and B is a European ISP, interconnecting in New York would 
introduce lower costs for A and higher costs for B than if they interconnect in Amsterdam. 
Similarly, the previous costs are route-dependent. 
For example, the cost of a transcontinental route is higher than the cost of a route that delivers
the traffic in the same metro area. 
Or, the cost of a route that goes through a transit provider is higher than the cost of a local
route. 

The exposition so far assumed that A exports a route to B and the latter sends the traffic volume 
$T$ through A. 
In practice, the interconnection will carry traffic in both directions. If we apply the Nash-Peering
framework separately to each direction of the interconnection, it could be that $A$ has to pay a fee $r_{AB}$ to B 
and $B$ has to pay a fee $r_{BA}$ to A. The two ASes can then combine the two payments into a single
lump-sum payment.

Settlement-free peering is a special case of Nash-Peering 
when $c'_B - c_B =  c'_A - c_A$, i.e., when the interconnection is equally beneficial to both ASes. 
Paid-peering, on the other hand, may seem initially as similar to Nash-Peering. This is not the 
case however for two reasons: first, in paid-peering it is always the sender (typically a content 
provider) that has to pay the receiver (typically an access provider), and second, the price associated
with paid-peering may not be determined based on the interconnection costs for the two ASes and  
the costs of their outside options.

We should emphasize that Nash-Peering does not consider the economic value of the information
that is carried by the traffic flow $T$ 
(e.g., whether the traffic is HBO premium content or whether it originates from Google). 
Considering that economic value and trying to somehow distribute it between the ASes that the traffic
goes through is a highly controversial subject and is viewed by many as a violation of network neutrality
\cite{economides2012network}. 
Nash-Peering does not do that -- it only considers the costs incurred by the two ASes in carrying 
this traffic through their infrastructure.

Equations (\ref{eq:utilityA}) and (\ref{eq:utilityB}) show that the utility of both ASes
increases with the surplus $\Delta$. 
So, given a certain outside option with costs $c'_A$ and $c'_B$, 
{\em both ASes have the same incentive}: to decrease their internal interconnection costs $c_A$ and $c_B$ 
as much as possible. 
As long as $(c_A+c_B)$ is minimized, both A and B will split the corresponding maximized surplus equally;
the relative magnitude of $c_A$ and $c_B$ does not matter.

What if the two ASes simplify their negotiations by considering many routes simultaneously, 
and averaging the underlying interconnection costs across all routes?\footnote{Each 
route would be weighted based on the traffic volume it carries.}  
This can be done in practice but it may result in loss of economic efficiency. 
To see why, consider a simple example with two routes.
Suppose that the surplus for route-1 is positive ($\Delta_1 = (c'_{A,1} + c'_{B,1}) - (c_{A,1}+c_{B,1})>0$) 
but the surplus for route-2 is negative. 
If the two ASes negotiate for both routes at the same time, averaging the corresponding cost 
parameters, it may be that the aggregate surplus $\Delta$ is positive.
In that case the interconnection will be established, even though both A and B would be better off 
if they had only interconnected for route-1.
Similarly, if $\Delta <0$, they will not interconnect even though they would be better off 
if they had only considered route-1.

\section{Disputes and Nash-Peering} \label{sec:disputes}
In this section we briefly review some common types of interconnection disputes 
(see \cite{bafna:2014} for a recent historical analysis) and 
discuss how they would be resolved in the Nash-Peering framework. 

\subsection{Access vs. Content providers}
Especially in the last few years,
there are frequent disputes between Access Providers (APs) and Content Providers (CPs).
The sf-Peering framework does not allow
these two types of ASes to negotiate their interconnection in a principled manner
based on actual costs and available alternatives.
APs and CPs are very different in terms of their traffic patterns:
the former mostly consume traffic while the latter mostly produce traffic,
meaning that  any peering conditions based on traffic ratios will obviously not be met.
CPs often request sf-Peering with APs (so that they both minimize upstream transit fees)
while APs claim that CPs benefit much more from such interconnections.

In the Nash-Peering framework, APs and CPs can always interconnect if there is a mutual benefit
for both of them (positive surplus). 
For instance, in an actual recent case between a major US ISP (Comcast) and a major video provider
(Netflix),
the CP's outside option was to pay one or more transit providers (or CDNs)
while the AP's outside option was to rely on one or more sf-Peering links with those transit providers,
 suggesting that $c'_{CP} \gg c'_{AP}$. 
Further, if the interconnection(s) between them would take place closer
to the CP's data centers, it would also be that $c_{AP} \gg c_{CP}$.
In a scenario like this, Nash-Peering specifies that the CP will have to pay the AP for the interconnection.
This payment may be close to zero, or it may even be reversed (from the AP to the CP), if
the CP has its own backbone network that delivers the traffic very close to the final 
``eyeballs'' (meaning that $c_{AP} \approx c_{CP}$ or even that $c_{AP} < c_{CP}$),
while the outside option is equally expensive for both of them (perhaps they both 
use the same transit provider, with $c'_{CP} \approx c'_{AP}$). 

\subsection{The fallacy of traffic ratios}
A second common dispute is between ISPs (transit or access providers) that have established  
sf-Peering interconnections based on a traffic ratio constraint $\gamma$ (i.e., A and B peer
if the traffic between them, $T_{AB}$ and $T_{BA}$, satisfies the constraint 
$1/\gamma < \frac{T_{AB}}{T_{BA}} < \gamma$, where $\gamma$ is typically between 2 and 5).
This condition, however, does not have have any relevance to the economic benefits of this
interconnection for each party.  
It is possible that the condition is {\em not} met, even though the interconnection's 
surplus is positive, meaning that the two ISPs would not establish an sf-Peering link even though
a Nash-Peering interconnection would be beneficial for both of them. 
The opposite can also happen, if the surplus is negative but the traffic ratio constraint is met.

The traffic ratio condition has its roots in the economics of telephone networks 
where every long-distance call would require roughly the same resources from the two carriers.
Nash-Peering replaces this constraint 
with economic considerations that do not make any assumptions about the correspondence between 
economic value or cost and the corresponding traffic flows.  

\subsection{Tier-1 ``peering wars''}
A third type of dispute arises due to the status of some ASes as ``Tier-1'' 
(meaning that they do not have any transit providers).  
In the current Internet, all Tier-1 providers have to be interconnected with a 
full-mesh of sf-Peering relations. The presence of such a clique results in a 
problematic state where existing Tier-1 providers adopt highly restrictive peering policies 
in order to avoid interconnecting with networks outside the clique (there is no economic incentive 
for a network in the clique to peer with a network outside the clique). 
Similarly, suppose that an AS X is a member of the Tier-1 clique but another AS Y claims 
that X no longer satisfies Y's peering criteria. If Y terminates the sf-Peering interconnection with X, 
some ASes in the customer cone of Y would be disconnected from some ASes in the customer cone of X. 
This risk of network partitioning causes some ASes to retain existing peering links even 
if they are no longer beneficial to them. 

In other words, the clique of Tier-1 providers is hard to evolve and its static structure can 
cause disputes between incumbents and newcomers. In the Nash-Peering framework, 
Tier-1 providers do not have any special privilege (other than not having an upstream transit provider). 
All their interconnections can be based on Nash-Peering, instead of sf-Peering, 
and there would be dynamic exchanges of money between them that change in terms of amount 
and direction from month to month. Further, a new entrant does not need to negotiate rigid and 
restrictive peering policies,  but can peer with any member of the clique 
(possibly with a payment) as long as it is mutually beneficial to do so. 

\subsection{Not all routes cost the same}
Finally, we believe that the current interconnection framework does not provide
the right incentives for ISPs to invest in high-cost routes that reach remote or
sparsely populated destinations such as rural regions. 
The main issue is that the same transit price is
typically applied on all routes, independent of the cost associated with each route.
At the same time, intense competition forces all transit providers to reduce their transit
prices as much as possible.
So, an ISP does not have the incentive to invest in the infrastructure that a high-cost route 
requires, given that that route will not generate additional revenue.  

With Nash-Peering, on the other hand, the interconnection fees are determined based on the cost of a route. 
If the cost $c_A$ of the route that A advertises to B increases, 
while the three other parameters of the interconnection ($c'_A, c'_B, c_B$) are constant, 
the payment $r$ from B to A will also increase (as long as $\Delta >0$). 
This gives A the incentive to invest in high-cost routes that would be under-provisioned today.

\section{Nash-peering in practice} \label{sec:practice}
So far, our exposition assumed that A and B will agree to adopt the Nash solution, without considering
the actual bargaining process between them or the information they will need during that process.  
Here, we explain how we envision Nash-Peering in a more practical context.

\subsection{The bargaining process}
A bargaining process
between two rational players, when modeled as a sequential game, converges to a unique subgame-perfect equilibrium
that is identical to the Nash solution \cite{binmore1986nash}. 
The mathematical assumptions for this result (see Proposition-3) 
are reasonable and we expect them to be true in practice 
(see Assumptions 1, 2, 8 and 9). The key condition is that the 
probability that the negotiations between A and B will fail, even though the surplus $\Delta$ is positive, 
is the same for both players and negligible. 
In other words, both players negotiate patiently and rationally, knowing that if they manage to
reach agreement they will mutually benefit compared to their outside option. 
Considering impatient or asymmetric players (e.g., A is an AS that desperately needs this interconnection, while 
B does not care much about it) is an important question for future research but outside the scope of this paper. 

Based on the previous result of \cite{binmore1986nash}, 
we can assume that the two ASes A and B will either directly agree to 
adopt the Nash solution, as a normative statement about how they should be sharing the costs of their
interconnection, or it is reasonable to expect that a bargaining process between them will eventually 
result in the Nash solution.  

\subsection{Interconnection parameters}
Even if A and B agree to adopt the Nash solution, what information will they need to have access to?
The Nash solution requires knowledge of $c_A$, $c_B$, $c'_A$, and $c'_B$.  
We refer to these four terms as the {\em interconnection parameters}:
they represent the costs that are associated with this traffic volume 
in the outside option ($c'_A$, $c'_B$), and with the corresponding 
costs if that interconection is actually established ($c_A$, $c_B$). 

First, an AS should be able to estimate the interconnection parameters for its own traffic.
The more challenging question is how to estimate the corresponding parameters of the other AS. 
Obviously, A should not just trust B when the latter claims that its relevant parameters 
are $c_B$ and $c'_B$. Later in this section, we describe an one-sided estimation process
that would allow A to estimate the parameters $c_B$ and $c'_B$ of AS B, at least approximately.  
More generally, however, we do not believe that these negotiations have to be based entirely on one-sided 
estimated parameters.
We envision that when two ASes negotiate their interconnection agreement they will be presenting quantitative
cost analyses to each other for the traffic they will be exchanging, allowing the other AS to verify 
claims about the cost of the outside option or of the negotiated interconnection.

\subsection{The granularity of negotiations}
In practice, the negotiations between A and B will not need to consider each BGP route separately.
First, even though there are about half a million BGP routes today, the vast majority of the traffic
that is directly exchanged between two ASes is dominated by a relatively small number of BGP routes 
\cite{labovitz2011internet}. 
For instance, if A is Comcast and B is Google, the two ASes could focus on the rather small number of 
routes that originate from Comcast's network,  given that the vast majority
of the traffic in this interconnection would flow from Google data centers to Comcast subscribers.   

Second, we expect that the interconnection parameters of large groups of routes will be roughly identical, 
when considering a specific location. 
For instance, if A is Cogent and B is Deutsche Telekom, and they consider interconnecting in Frankfurt,
most Cogent routes for US-based destination networks would probably be grouped together.
On the other hand, Cogent routes for Asia-based destination networks would probably be associated with
different interconnection parameters, and they would be negotiated as a different group of routes.
Also, these parameters would be different when Cogent and Deutsche Telekom 
interconnect in Ashford VA, instead of Frankfurt.

A quantitative analysis of how many groups of routes would need to be negotiated separately, at a given 
interconnection location and for a given pair of ASes, is certainly an important question 
that we will investigate in future work. 
It is clear though that negotiating large bundles of routes with similar interconnection parameters
will significantly decrease the accounting complexity of Nash-Peering.

\subsection{One-sided parameter estimation}
We assume that a network with complete knowledge of its internal infrastructure
can estimate its own internal costs (interconnect and backhaul), 
perhaps as proposed by Motiwala et al.~\cite{motiwala:2012}.

An AS also needs to estimate
the internal costs and outside options of its potential peers. 
To estimate the outside option of B, the first option is for A to use BGP and traceroute measurements
from a route server in B's network, along with AS-relationship data~\cite{caida-asrank}, 
to infer whether B's path toward destination D is via a customer, provider, or peer. 
A second option is to use {\em AS customer cone data}~\cite{caida-asrank} to 
infer whether B could reach prefix D using either customer or peer links. 
If B's outside option is a provider, A can estimate
B's transit cost using available transit pricing data.

To estimate B's internal costs, A needs to know the ingress and egress points
into B's network for the traffic toward destination D. 
A can use traceroutes from looking glass servers in 
B (or networks in B's customer cone), or the reverse traceroute
system~\cite{bassett:2010}, to measure the path towards D, 
and hence infer the ingress and egress points into B's network. 
Assuming symmetry of cost factors, A can approximate B's internal
costs using its own cost model and the ingress/egress points into B's network.

\section{Related work} \label{sec:related}
Dhamdhere et al.~\cite{dhamdhere:2010} proposed a {\em value-based peering scheme} 
that treats each interconnection as a paid-peering link, where a payment is exchanged depending on
the ``added value'' that the interconnection offers to each party. 
Courcoubetis et al.~\cite{courcoubetis:2016} have also applied the theory of Nash bargaining 
to derive paid-peering prices (specifically targeting the negotiation between content and
access providers) considering factors such as advertising revenues, 
subscriber loyalty, and interconnection or capacity costs. 
Jahn and Pr{\"u}fer~\cite{jahn2008interconnection} analyzed a model in which two ISPs with 
asymmetric sizes compete for subscribers 
while at the same time they consider paid-peering, based on Nash bargaining, as a way to interconnect. 
They assume equal internal costs for both ISPs 
and a variable transit fee charged by a single transit provider.
Besen et al.~\cite{besen2001advances} analyzed the outcome of Nash bargaining 
between two peers in the absence of an outside option, 
which incurs service degradation, and loss of customers and revenue. 
Nash-Peering differs from these earlier models in that the interconnection fees are determined based on costs
rather than the economic ``value'' of a flow (which is much harder to measure -- and rather controversial).
 
Valancius et al.~\cite{valancius:2011} demonstrated that ISPs could
maximize transit profits by employing a {\em tiered pricing} scheme, 
charging different prices for different routes -- they  
also showed that a small number of tiers would be sufficient. 
Their scheme focuses on transit relations, however, and does not involve price bargaining. 
MINT~\cite{valancius:2008} and Route Bazaar~\cite{castro-hotos:2015} are two 
schemes where ``route sellers'' advertise connectivity for different {\em path segments} 
and ``route buyers'' compose end-to-end paths by purchasing connectivity over specific path segments. 
Route Bazaar and MINT do not focus on the economics of how the prices are set, 
or how prices relate to costs and outside options. 
Tan et al. ~\cite{tan2006economic} showed how providers can establish interconnection agreements 
to dynamically trade network capacity. This demonstrates how ``paid peering" can be employed to 
improve network utilization and create incentives to improve infrastructure. 

Prior research by Ma et al.~\cite{ma-shapley:2010,ma-cooperative:2011} employs
cooperative game theory to model multilateral interconnections betweeen
different types of networks in the Internet ecosystem. They proposed
the use of the {\em Shapley Value} to achieve fair and stable division
of surplus among different types of providers. Nash-Peering differs from that approach in that
it only requires bilateral contracts between ASes. Multilateral interconnection
agreements are much harder to establish and manage in practice.  

From a policy standpoint, Faratin et al.~\cite{faratin:08} discuss the
emergence of paid peering, highlight the increased complexity of the
negotiations involved in peering, and discuss the implications for
telecommunications policy. Clark et al.~\cite{clark:2011} 
analyzed the reasons why settlement-free peering
based on traffic ratios could make way for paid-peering between CDNs
and access ISPs. The authors advocated increased transparency into
costs, traffic patterns, and interconnection terms over regulation of interconnection. 
Economides and T{\aa}g~\cite{economides2012network} focus on network neutrality regulation and 
demonstrate the effects of regulation on the total surplus of peering relations.

\section{Discussion} \label{sec:discussion}
Nash-Peering represents a major departure from the current interconnection framework.
Instead of the limiting dichotomy of AS relations into transit and sf-Peering, 
it offers a more general, efficient and fair framework in which two ASes evaluate the economic benefits that an 
interconnection would provide them, at the level of groups of routes with similar costs, 
and split any potential surplus equally among them.

The next step in this research will be to examine 
the Internet-wide effects of bilateral Nash-Peering interconnections.
For instance, a change in the interconnection between A and B may trigger changes in many  
interdomain routes. The latter may affect the traffic volume in other interconnections,
between different pairs of ASes, i.e., these bilateral interconnections have network-wide externalities. 
Of course such externalities are also present in the current interconnection framework 
but Nash-Peering will probably introduce more complex dynamics due to the finer granularity of 
bilateral negotiations.
Interestingly, the network externalities of bilateral contracts is also 
an active research area in economic theory, and we plan to leverage that emerging body of knowledge 
\cite{fontenay2014bilateral}.

\newpage

\bibliographystyle{abbrv}
\bibliography{refs}

\begin{thebibliography}{10}

\bibitem{bafna:2014}
S.~Bafna, A.~Pandey, and K.~Verma.
\newblock {Anatomy of Internet Peering Disputes}.
\newblock {\em CoRR}, abs/1409.6526, 2014.

\bibitem{besen2001advances}
S.~Besen, P.~Milgrom, B.~Mitchell, and P.~Srinagesh.
\newblock {Advances in routing technologies and Internet peering agreements}.
\newblock {\em The American Economic Review}, 91(2):292--296, 2001.

\bibitem{binmore1986nash}
K.~Binmore, A.~Rubinstein, and A.~Wolinsky.
\newblock {The Nash bargaining solution in economic modelling}.
\newblock {\em The RAND Journal of Economics}, pages 176--188, 1986.

\bibitem{castro-hotos:2015}
I.~Castro, A.~Panda, B.~Raghavan, S.~Shenker, and S.~Gorinsky.
\newblock {Route Bazaar: Automatic Interdomain Contract Negotiation}.
\newblock In {\em Proceedings of the 15th USENIX Conference on Hot Topics in
  Operating Systems}, HOTOS'15. USENIX Association, 2015.

\bibitem{caida-asrank}
{Center for Applied Internet Data Analysis}.
\newblock {CAIDA - AS Rank}.
\newblock \url{http://as-rank.caida.org}.

\bibitem{clark:2011}
D.~Clark, W.~Lehr, and S.~Bauer.
\newblock {Interconnection in the Intenret: The Policy Challenge}.
\newblock In {\em Telecommunications Policy Research Conference (TPRC)}, Aug
  2011.

\bibitem{clark2014measurement}
D.~D. Clark, S.~Bauer, W.~Lehr, K.~Claffy, A.~D. Dhamdhere, B.~Huffaker, and
  M.~Luckie.
\newblock {Measurement and analysis of Internet interconnection and
  congestion}.
\newblock In {\em 2014 TPRC Conference Paper}, 2014.

\bibitem{courcoubetis:2016}
C.~Courcoubetis, L.~Gyarmati, N.~Laoutaris, P.~Rodriguez, and K.~Sdrolias.
\newblock Negotiating premium peering prices: A quantitative model with
  applications.
\newblock {\em ACM Trans. Internet Technol.}, 16(2):14:1--14:22, Apr. 2016.

\bibitem{dhamdhere:2010}
A.~Dhamdhere, C.~Dovrolis, and P.~Francois.
\newblock {A Value-based Framework for Internet Peering Agreements}.
\newblock In {\em International Teletraffic Congress (ITC)}, Amsterdam, The
  Netherlands, Oct 2010.

\bibitem{economides2012network}
N.~Economides and J.~T{\aa}g.
\newblock {Network neutrality on the Internet: A two-sided market analysis}.
\newblock {\em Information Economics and Policy}, 24(2):91--104, 2012.

\bibitem{faratin:08}
P.~Faratin, D.~Clark, S.~Bauer, W.~Lehr, P.~Gilmore, and A.~Berger.
\newblock {The Growing Complexity of Internet Interconnection}.
\newblock {\em Communications \& Strategies}, 1(72):51--72, 2008.

\bibitem{fcc15}
{Federal Communications Commission}.
\newblock {Open Internet Order}, 2015.

\bibitem{fontenay2014bilateral}
C.~C. Fontenay and J.~S. Gans.
\newblock Bilateral bargaining with externalities.
\newblock {\em The Journal of Industrial Economics}, 62(4):756--788, 2014.

\bibitem{huston:1998}
G.~Huston.
\newblock {\em ISP Survival Guide: Strategies for Running a Competitive ISP}.
\newblock John Wiley \& Sons, Inc., New York, NY, USA, 1999.

\bibitem{jahn2008interconnection}
E.~Jahn and J.~Pr{\"u}fer.
\newblock {Interconnection and competition among asymmetric networks in the
  Internet backbone market}.
\newblock {\em Information Economics and Policy}, 20(3):243--256, 2008.

\bibitem{bassett:2010}
E.~Katz-Bassett, H.~V. Madhyastha, V.~K. Adhikari, C.~Scott, J.~Sherry,
  P.~Van~Wesep, T.~Anderson, and A.~Krishnamurthy.
\newblock Reverse traceroute.
\newblock In {\em Proceedings of the 7th USENIX Conference on Networked Systems
  Design and Implementation}, NSDI'10, pages 15--15. USENIX Association, 2010.

\bibitem{labovitz2011internet}
C.~Labovitz, S.~Iekel-Johnson, D.~McPherson, J.~Oberheide, and F.~Jahanian.
\newblock Internet inter-domain traffic.
\newblock {\em ACM SIGCOMM Computer Communication Review}, 41(4):75--86, 2011.

\bibitem{lodhi:2014}
A.~Lodhi, N.~Larson, A.~Dhamdhere, C.~Dovrolis, and k.~claffy.
\newblock {Using peeringDB to Understand the Peering Ecosystem}.
\newblock {\em SIGCOMM Comput. Commun. Rev.}, 44(2):20--27, Apr. 2014.

\bibitem{ma-shapley:2010}
R.~T.~B. Ma, D.~M. Chiu, J.~C.~S. Lui, V.~Misra, and D.~Rubenstein.
\newblock {Internet Economics: The Use of Shapley Value for ISP Settlement}.
\newblock {\em IEEE/ACM Trans. Netw.}, 18(3):775--787, June 2010.

\bibitem{ma-cooperative:2011}
R.~T.~B. Ma, D.~M. Chiu, J.~C.~S. Lui, V.~Misra, and D.~Rubenstein.
\newblock {On Cooperative Settlement Between Content, Transit, and Eyeball
  Internet Service Providers}.
\newblock {\em IEEE/ACM Trans. Netw.}, 19(3):802--815, June 2011.

\bibitem{motiwala:2012}
M.~Motiwala, A.~Dhamdhere, N.~Feamster, and A.~Lakhina.
\newblock Towards a cost model for network traffic.
\newblock {\em SIGCOMM Comput. Commun. Rev.}, 42(1):54--60, Jan. 2012.

\bibitem{nash1950bargaining}
J.~F. Nash~Jr.
\newblock The bargaining problem.
\newblock {\em Econometrica: Journal of the Econometric Society}, pages
  155--162, 1950.

\bibitem{nash1953two}
J.~F. Nash~Jr.
\newblock Two-person cooperative games.
\newblock {\em Econometrica: Journal of the Econometric Society}, pages
  128--140, 1953.

\bibitem{tan2006economic}
Y.~Tan, I.~R. Chiang, and V.~S. Mookerjee.
\newblock {An economic analysis of interconnection arrangements between
  Internet backbone providers}.
\newblock {\em Operations research}, 54(4):776--788, 2006.

\bibitem{valancius:2008}
V.~Valancius, N.~Feamster, R.~Johari, and V.~Vazirani.
\newblock {MINT: A Market for INternet Transit}.
\newblock In {\em Proceedings of the 2008 ACM CoNEXT Conference}, CoNEXT '08,
  pages 70:1--70:6. ACM, 2008.

\bibitem{valancius:2011}
V.~Valancius, C.~Lumezanu, N.~Feamster, R.~Johari, and V.~V. Vazirani.
\newblock How many tiers?: Pricing in the internet transit market.
\newblock In {\em Proceedings of the ACM SIGCOMM 2011 Conference}, SIGCOMM '11,
  pages 194--205. ACM, 2011.

\bibitem{norton-peering-playbook}
{William B. Norton}.
\newblock {\em {The Internet Peering Playbook: Connecting to the Core of the
  Internet}}.
\newblock {DrPeering Press}, 2012.

\end{thebibliography}

\end{document}